# Recent Progresses and Perspectives of UV Laser Annealing Technologies for Advanced CMOS Devices

Toshiyuki Tabata *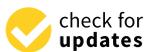, Fabien Rozé, Louis Thuries, Sébastien Halty, Pierre-Edouard Raynal, Imen Karmous and Karim Huet

Laser Systems & Solutions of Europe (LASSE), 145 Rue Des Caboeufs, 92230 Gennevilliers, France
\* Correspondence: toshiyuki.tabata@screen-lasse.com

**Abstract:** The state-of-the-art CMOS technology has started to adopt three-dimensional (3D) integration approaches, enabling continuous chip density increment and performance improvement, while alleviating difficulties encountered in traditional planar scaling. This new device architecture, in addition to the efforts required for extracting the best material properties, imposes a challenge of reducing the thermal budget of processes to be applied everywhere in CMOS devices, so that conventional processes must be replaced without any compromise to device performance. Ultraviolet laser annealing (UV-LA) is then of prime importance to address such a requirement. First, the strongly limited absorption of UV light into materials allows surface-localized heat source generation. Second, the process timescale typically ranging from nanoseconds (ns) to microseconds (μs) efficiently restricts the heat diffusion in the vertical direction. In a given 3D stack, these specific features allow the actual process temperature to be elevated in the top-tier layer without introducing any drawback in the bottom-tier one. In addition, short-timescale UV-LA may have some advantages in materials engineering, enabling the nonequilibrium control of certain phenomenon such as crystallization, dopant activation, and diffusion. This paper reviews recent progress reported about the application of short-timescale UV-LA to different stages of CMOS integration, highlighting its potential of being a key enabler for next generation 3D-integrated CMOS devices.

## 1. Introduction

Nowadays, sustainable, high-performance, and energy-efficient nanoelectronics drive the further development of CMOS technologies, not only by traditional planar scaling, called "More Moore", but also by opening the integration dimension towards the third axis perpendicular to the silicon (Si) wafer plane. This three-dimensionally (3D) stacked device architecture alleviates the difficulties possibly encountered in the way of pursuing the traditional planar scaling of transistors, while allowing the continuous increment of effective chip density and performance thereby. Here, a "monolithic" or "sequential" approach is talked about, but not a "packaging" one, where interlayer connectivity is dominated by chip bonding alignment accuracy. It also brings another benefit of implementing additional functionalities in CMOS devices, being called "More Than Moore". Furthermore, over the coming end of the current CMOS scaling, "Beyond CMOS" is preconized to determine the best means for continuing technological advancements, having new materials, concepts, and architectures.

Among these three axes of evolution, "More Than Moore" would be the one which has been the most investigated over the last several years, in parallel with the evolution of ultra-violet laser annealing (UV-LA) technologies in the semiconductor industry. To realize 3D-integrated CMOS devices, a new electrically functional Si (or other semiconductor materials) layer must be fabricated directly on the underlayer components, either by wafer bonding [1] or by deposition [2]. Of course, it also requires other steps. For instance, to form a source and drain (S/D), recrystallization and dopant activation are necessary after ion implantation. Otherwise, epitaxy enables in situ doping with good crystal quality. To form a gate stack, dielectric and work function metal deposition, sometimes followed by reliability annealing, are required. Some of them typically need high-temperature processing. Those steps must be accomplished by reducing the thermal budget, which is empirically discussed as a simple combination of temperature (i.e., activation energy) and time (i.e., kinetics) effects, while avoiding any drawback in terms of the performance of each module. Today, as shown in Figure 1, the thermal budget acceptable for 3D-integrated CMOS devices is known as 500 °C for a few hours [1,3,4], and might be further reduced in future developments. UV-LA is then of prime importance to address such requirements for the following reasons. First, the use of a UV light enables strongly limited absorption into materials, and thereby surface-localized heat source generation. Second, the UV-LA process timescale, typically ranging from nanoseconds (ns) to microseconds (µs), efficiently prevents the generated heat from diffusing deeply in the vertical direction. In a given 3D stack, these specific features allow the actual process temperature to be elevated in the top-tier layer without introducing any drawback in the bottom-tier one. In fact, although UV-LA itself has existed for decades, most of the studies of those precedent eras were conducted in a laboratory-scale experiment but not in real industrial-quality devices. This is certainly because there was no UV-LA equipment ready to be implemented in high-volume manufacturing. On the other hand, in the current era, our 300 mm process-compatible UV-LA equipment is commercialized [5], with the capability of UV-LA process simulation [6] to support the UV-LA integration into existing device fabrication flows.

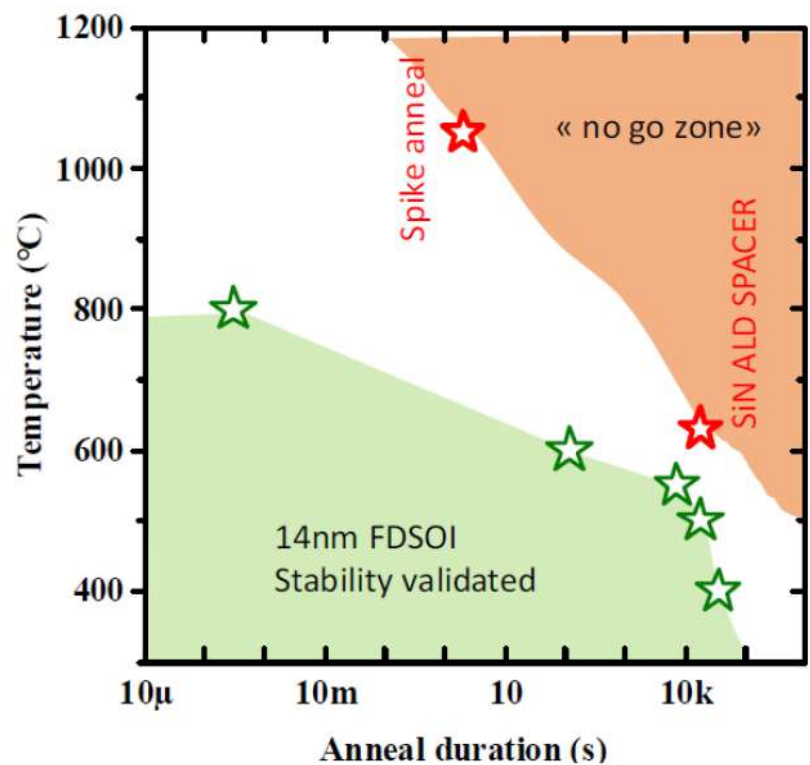

**Figure 1.** Acceptable thermal budget to preserve the performance of bottom-tier devices. Reprinted with permission from Ref. [3]. 2017, IEEE.

Considering the large freedom of designing applications in 3D integration (e.g., logic-on-logic [3], memory-on-logic [7–9], and sensor-on-logic [3,9]), the thermal budget would have to be managed application-by-application. Furthermore, in some cases components are inserted into interconnect (i.e., the back-end-of-line (BEOL)) layers to enhance the benefits of energy efficiency and look to future neuromorphic computing [10–13]. In such a context, a clear need to freely control the thermal budget is rising. An interesting study is reported in Ref. [14], where the impacts of the buried-oxide (i.e., a Si dioxide ($SiO_2$) layer separating the top and bottom layers in a 3D stack) thickness and applied thermal budget on the dopant activation in an ion-implanted top and bottom Si layers (i.e., a Si substrate as the bottom layer, whereas an amorphous Si thin layer as the top layer) are systematically investigated. It is noteworthy that a thicker $SiO_2$ interlayer results in more efficient thermal isolation. Moreover, with a given UV light, laser fluence can be another knob of thermal isolation management. As shown in Figure 2, a more realistic study is conducted using a 28 nm fully depleted Si-on-insulator (FDSOI) CMOS integration, where the impacts of UV-LA on underlying components (i.e., bottom MOS devices and BEOL) are assessed in terms of the performance of transistors and copper (Cu)-based interconnects [4]. The results show no significant degradation in both, confirming the advantage and thermal budget compatibility of UV-LA in 3D integration.

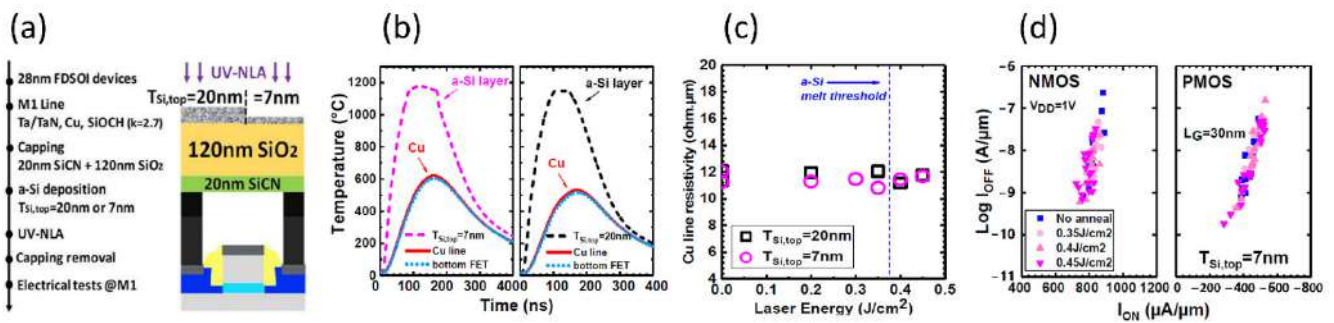

**Figure 2.** (**a**) Process flow and schematic figure of the 3D structure treated by UV-LA, (**b**) Simulated time–temperature profiles of the applied UV-LA processes, (**c**) Cu line resistivity evaluated before and after UV-LA, (**d**) $I_{ON}$-$I_{OFF}$ distribution of the bottom n- and p-type MOS devices before and after UV-LA. Reprinted/adapted with permission from Ref. [4]. 2020 IEEE.

In the following sections, we aim at highlighting the potential benefits of short-timescale UV-LA in materials engineering, crossing over diverse applications from the front-end-of-line (FEOL) to BEOL.

## 2. FEOL Applications

### 2.1. Reliability Annealing for High-k/SiO$_2$/Si Gate Stacks

In recent planar transistors (e.g., 7 nm node CMOS technology, as shown in Ref. [15]), the $SiO_2$/Si interface is still utilized in the gate stack (generally a stack of a few nm thick high-permittivity (high-*k*) hafnium dioxide ($HfO_2$) film on a Si substrate with a subnanometer-thick $SiO_2$ interlayer), requiring a lot of efforts to further scale the equivalent oxide thickness (EOT) in a subnanometer region. This gate stack needs to be exposed to a high-temperature annealing process, typically ranging from 800 °C to 1000 °C, for a few seconds in order to ensure temperature instability (BTI) improvement [16–18].

The origin of the BTI is the electrical traps existing at the gate stack interface and within the dielectric films [19]. In the $HfO_2$/$SiO_2$/Si stack, there are different BTI sources. One of them is $HfO_2$ bulk traps, and it is proposed to insert a dipole between the $HfO_2$ and $SiO_2$ layers [18] so that the Si channel does not electrically communicate with such defect levels in device operation. Another source is the traps within the interfacial $SiO_2$/Si system (i.e., the bulk of the thin film and interface), where $SiO_2$ is often chemically grown to have a very thin film thickness (i.e., ~1 nm or less). Although annealing is necessary to cure or passivate those electrical traps, it must comply with the thermal budget limitations. Recently, low-temperature (from 100 °C to 450 °C) hydrogen-based annealing [20] and plasma treatment [21] have been proposed as an alternative to the conventional high-temperature annealing.

In fact, short timescale UV-LA may also address this problem. A chemically grown $SiO_2$ thin film is known to have a smaller density than thermally grown films, as well as some impurities remaining inside [22]. Such a situation would also be the case for a subnanometer-thick $SiO_2$ thin film grown by wet chemical cleaning for a CMOS gate stack. High-temperature processing enabled by UV-LA is then expected to help the removal of the remaining impurities in parallel with the nonequilibrium atomic bonding rearrangement possibly occurring thanks to its short timescale. We have conducted a preliminary study [23] relevant to the potential BTI source annihilation in the wet chemically grown $SiO_2/Si$ system by ns UV-LA (Figure 3), where the effective process temperature is set at 900 °C and the dwell time is in the order of $10^{-7}$ s. Interestingly, by accumulating the applied ns UV-LA process up to 1000 times, the rearrangement of the Si-O-Si network is observed by attenuated total reflectance (ATR) Fourier transform infrared (FTIR) spectroscopy and X-ray photoelectron spectroscopy (XPS). This coincides with the evolution of the electrical traps captured by a temperature-scan deep-level transient spectroscopy (DLTS). Then, the $P_b$ center-type traps seem annihilated, but it competes with the formation of another type of trap, so-called "tail states" or "U-shaped continuum", which could be associated to Si–O weak bonds formed at the $SiO_2/Si$ interface. It is therefore supposed that UV-LA alone may not be sufficient to perfectly build up the $SiO_2/Si$ system, but a potential of the nonequilibrium $SiO_2/Si$ system engineering is clearly demonstrated.

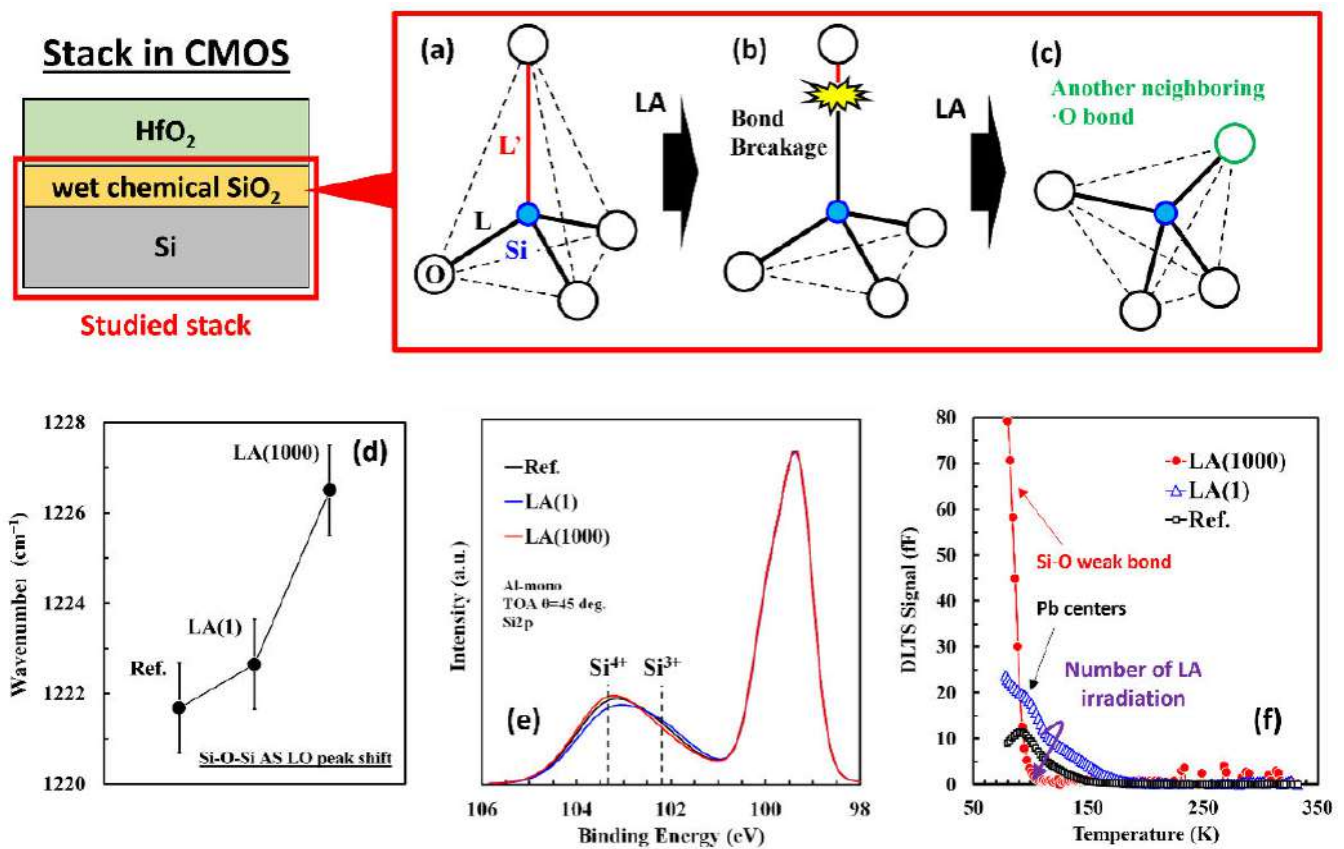

**Figure 3.** (**a**–**c**) Schematic figures of a possible local rearrangement procedure of the tetrahedral $SiO_4$ network. (**d**) Shift of the longitudinal optic (LO) phonon peak of the Si–O–Si asymmetric stretching (AS) vibration as a function of the number of LA irradiation, where LA(1) and LA(1000) stand for 1 and 1000 times irradiation, respectively. (**e**) Normalized Si 2p XPS spectra of the as-grown (Ref.), LA(1), and LA(1000) samples. (**f**) Comparison of the DLTS signals taken for the as-grown (Ref.), LA(1), and LA(1000) samples. Reprinted/adapted with permission from Ref. [23]. 2020, The Japan Society of Applied Physics.

*2.2. Channel Doping Engineering to Mitigate Short Channel Effects*

The short channel effects (SCE) are known as one of the most critical issues when scaling CMOS devices [24]. To alleviate the SCEs, shallow junction has been conceived, introducing source and drain extension doping beside the channel [25]. Therefore, the control of dopant diffusion in this extension area is critical for device performance. To that end, ns UV-LA may be an ideal solution because, in such a short timescale, dopant diffusion would be strongly limited (or even negligible). Indeed, for conventional dopants such as boron (B), phosphorus (P), and arsenic (As) in Si, it is the case under the Si melting point (i.e., ns UV-LA process does not melt the Si substrate) [26]. When melting the doped Si substrate, although the diffusivity of the dopants strongly enhances (typically towards the order of $10^{-4}$ cm$^2$ s$^{-1}$, as reported in Ref. [27]), their diffusion is limited within the melted region, having a box-like profile [28–31].

A practical example taken in real devices is reported in our previous work [32], where top planar FDSOI transistors having an 11 nm thick Si channel were fabricated (see Figure 4a for their process flow). After the ion implantation in the extension part, ns UV-LA was applied to fully melt the from 5 to 7 nm thick amorphized Si layer (Figure 4b,c) that regrows into the monocrystalline state (Figure 4d,e). After this Si channel regrowth, a raised S/D is properly formed by epitaxy, indicating that ns UV-LA does not seriously degrade the surface morphology of the regrown Si thin layer. In terms of the impact on the SCE, ns UV-LA clearly shows a benefit for suppressing the drain-induced barrier lowering (DIBL) compared to high-temperature spike annealing (Figure 4f), especially with the dopants such as B and P, which are easier to diffuse in Si than As [26].

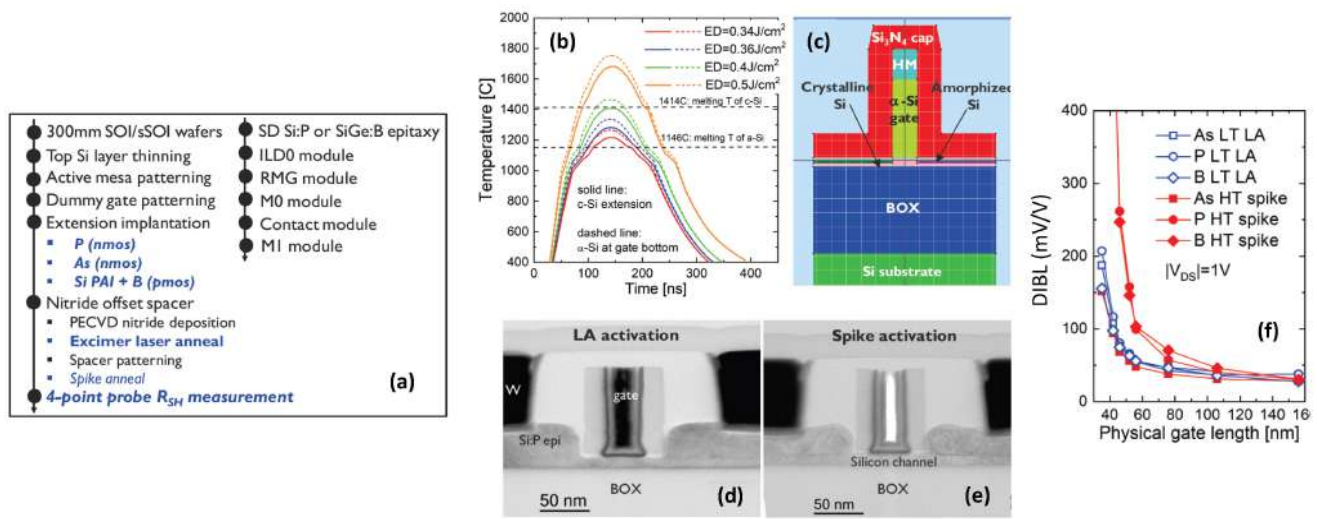

**Figure 4.** (**a**) Process flow of the FDSOI transistors treated by UV-LA; (**b**) Simulated time–temperature profiles of the applied UV-LA processes; (**c**) Structure used for the simulation; (**d**,**e**) Cross-sectional transmission electron microscopy (TEM) images of the EDSOI transistors after UV-LA and spike annealing; (**f**) DIBL extracted for different physical gate lengths. Reprinted/adapted with permission from Ref. [32]. 2020, IEEE.

### 3. MOL Applications

The word "middle-of-line (MOL)" stands for the set of processes relevant to the formation of the electrical connection of transistors prior to the BEOL modules. Especially, the low resistive "contact" formation has become one of the biggest challenges in recent years because it is a dominant parasitic factor in scaled transistors. This is a Schottky contact formed by a metal–semiconductor interface. Therefore, to reduce the resistivity, increasing the active carrier concentration in the semiconductor side is supposed to be a strategy of choice. In fact, the required level of specific contact resistivity for future nodes is going to be less than $1 \times 10^{-9}$ Ω cm$^2$ [33,34]. To achieve such a low resistivity of the contact, the active carrier concentration needs to be close to $1 \times 10^{21}$ at./cm$^3$ [35,36], or even higher [37]. Then, the carrier transport at the Schottky interface should be dominated by the field-emission model [38].

From the viewpoint of the UV-LA processes, there are two approaches. The first one is the liquid phase epitaxial regrowth (LPER), where the doped semiconductor is once melted and epitaxially regrown while activating the dopants. Typically, ns UV-LA enables this approach [34,39–43]. In this article, when discussing LPER, we have a regime so-called "Secondary Melting (SM)" in mind rather than the one so-called "Explosive Melting (EM)". In SM, the melted (i.e., liquid) layer forms at the surface of the semiconductor substrate and simply extends downward [31,44]. On the other hand, in EM, a thin liquid layer (typically a few to 10 nm of thickness [45]) formed at the surface travels in the depth direction, inducing recrystallization into a polycrystalline state [31,44]. The LPER-induced rapid solidification leads to the metastable incorporation of the dopants into the regrown semiconductor crystal. The melting of a doped semiconductor substrate may result in the degradation of the regrown surface morphology [1,40,42], and subsequent steps in a transistor fabrication flow might suffer from it. The second approach is the solid phase epitaxial regrowth (SPER), where the semiconductor layer amorphized by the ion implantation of the dopants can be regrown into the monocrystalline state without melting. Although ns UV-LA can achieve SPER and metastable dopant activation thereby, it requires multiple processes (i.e., more than 20 times the irradiation of the laser pulse) [46]. Therefore, in terms of the productivity, ns UV-LA might not be an optimal option. To address it, extending the process timescale towards µs scale may help [47]. In the previously reported ns UV-LA SPER, the surface morphology does not show serious degradation [46].

### 3.1. Dopant Activation by Liquid Phase Epitaxial Regrowth (LPER)

A potential advantage of LPER on contact resistivity lowering is that, if a doping element is properly selected, its segregation towards the surface occurs during the solidification of the melted semiconductor material. This may enhance the active carrier concentration near the metal/semiconductor interface.

For that, the segregation coefficient of a doping element ($k = C_S/C_L$, where $C_S$ and $C_L$ stand for the dopant concentrations in the solid (*s*) and liquid (*l*) phases at the vicinity of the moving $l/s$ interface) must be less than the unity (i.e., $k < 1$). It should be noted that this $k$ value is a function of the solidification front velocity ($V$) as shown in Figure 5 [48,49], and increasing $V$ makes $k$ become closer to the unity. For instance, antimony (Sb) in Si (i.e., n-type contact) shows such a surface segregation during ns UV-LA-induced LPER [42], and gallium (Ga) [39–41,43], aluminum (Al) [41], and indium (In) [41] in SiGe (i.e., p-type contact) also do (see Figure 6a,b) as experimental examples.

On the other hand, the activation of these segregated dopants seems not so simple. Firstly, when comparing the ratio of the LPER-induced active carrier concentration to the solid solubility limit of each doping case, that of Al in SiGe is almost unity at different melt conditions. This might be related to the self-compensation of the Al atoms (i.e., electrical deactivation due to lack of a bond between them when their doping concentration becomes extremely high) [50]. Therefore, not every small $k$ dopant may work. Secondly, as shown in Figure 7, the activation also depends on $V$, drawing a downward convex shape as an overall trend given by the whole red and blue data points. In fact, there is an interesting theoretical prediction about the minimum substitutional concentration (of In in Si) as a function of $V$ [51]. If the nonequilibrium feature of the segregation coefficient ($k$) is considered (i.e., $k = f(V)$), the relation between this concentration and $V$ is also drawn by a downward convex shape (Figure 8). Hence, increasing $V$ is suggested to enhance the active carrier concentration in LPER. However, it should be noted that it is in a trade-off with the efficiency of surface segregation (i.e., $k$ becomes closer to the unity when increasing $V$), and a compromise would have to be found when integrating ns UV-LA LPER into real CMOS contact modules.

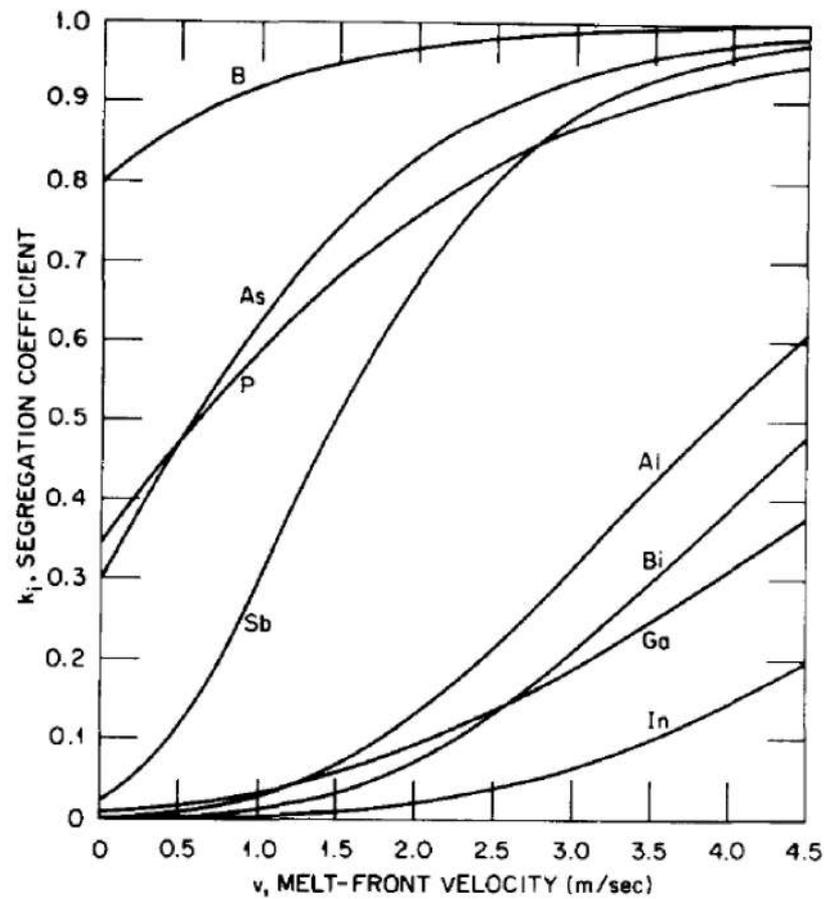

**Figure 5.** An example of the theoretically predicted dependence of the segregation coefficient ($k_i$) of dopants in Si on the solidification front velocity (i.e., the melt–front velocity, $v$). Reprinted with permission from Ref. [48]. 1980, AIP Publishing.

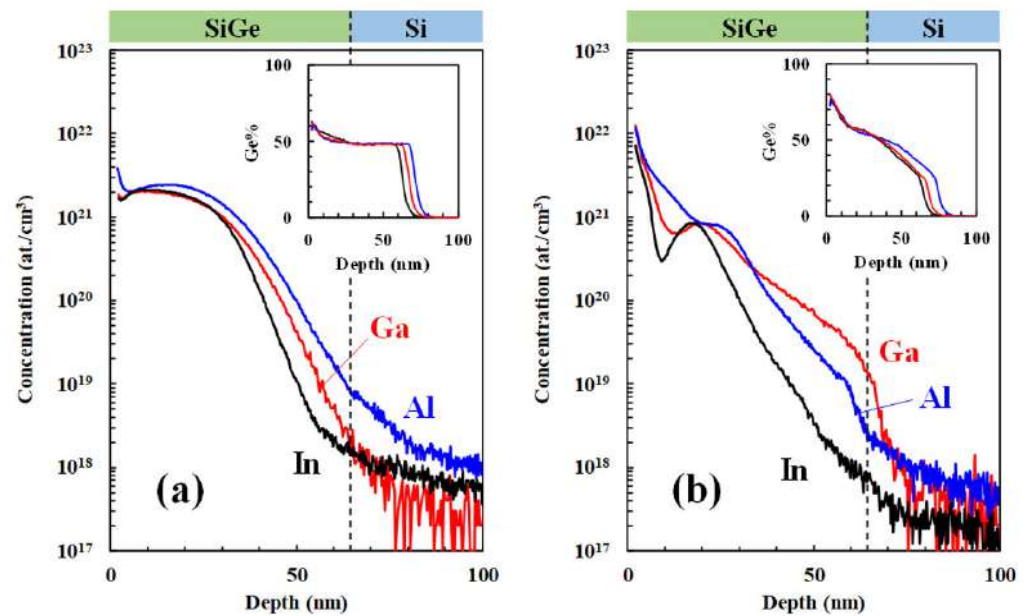

**Figure 6.** Secondary ion mass spectrometry (SIMS) profiles of Ga, In, and Al taken in the (**a**) as-implanted and (**b**) ns UV-LA-induced full SiGe epilayer melt samples. Reprinted with permission from Ref. [41]. 2019, The Japan Society of Applied Physics.

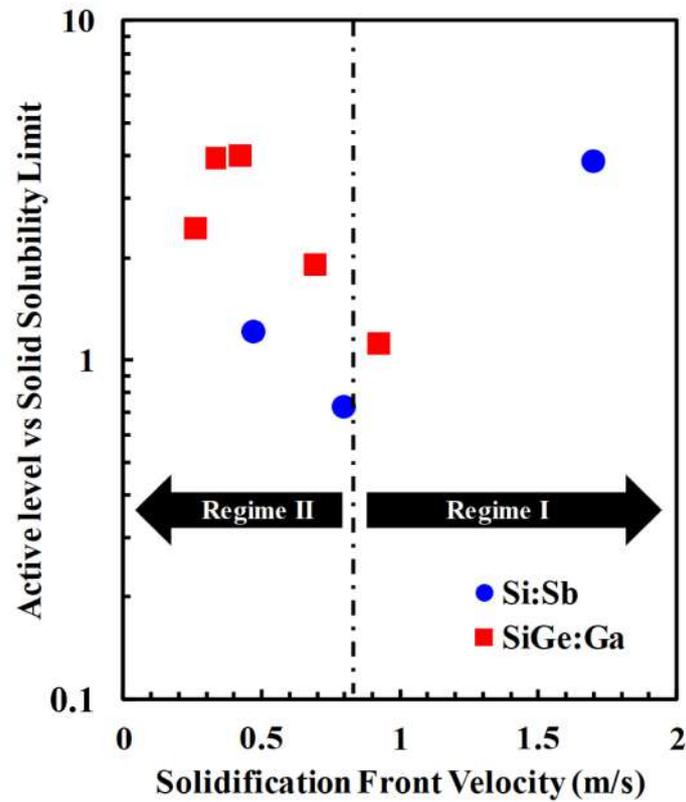

**Figure 7.** Plots of the degree of surpassing the solid solubility limit as a function of the simulated solidification front velocity (*V*) for the Sb-implanted Si and Ga-implanted SiGe samples. Reprinted with permission from Ref. [42]. 2020, AIP Publishing.

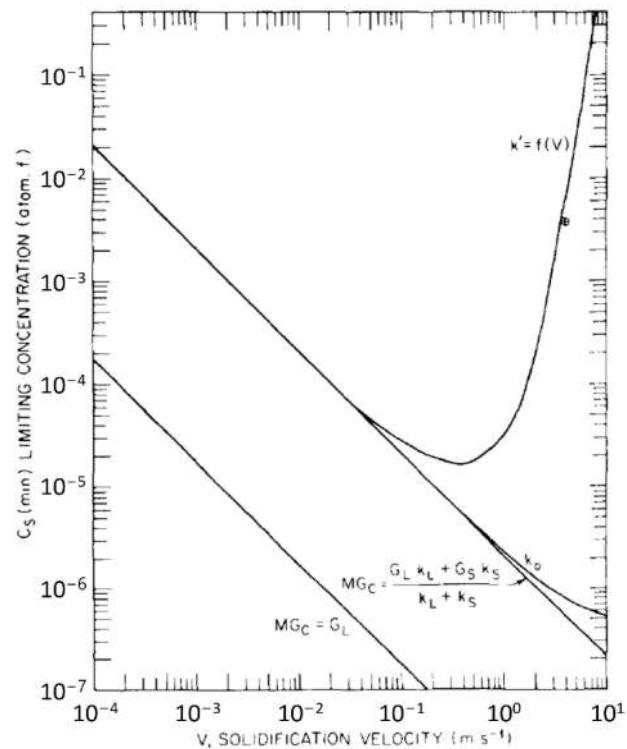

**Figure 8.** Theoretical prediction of the minimum substitutional concentration (of In in Si) as a function of *V*. The downward convex shape curve represents the case considering the nonequilibrium feature of the segregation coefficient. Reprinted/adapted with permission from Ref. [51]. 1981, AIP Publishing.

Recently, we have accessed it by using Sb doping for 14 nm node generation FinFET's Si-based contact (Figure 9) [52]. Although there is no electrical data, the segregation of ion-implanted Sb atoms at the top of the fin structure is clearly evidenced (Figure 9d,e, see the position indicated by the arrow "S/D epi top surface"). The *V* extracted by 3D TCAD simulation is about 4 m/s (Figure 9c). In fact, in real devices, the volume of the S/D parts is negligible compared to the underlying Si substrate, so that heat dissipation becomes fast, and V increases thereby. This value is in a promising range to have a high active carrier concentration, considering the Sb solid solubility limit in Si ($\sim 6.8 \times 10^{19}$ at./cm$^3$ [53]) and the expected gain from Figure 7 (more than 10 times at 4 m/s). Furthermore, other promising results are obtained in the p-type contacts of planer FDSOI transistors, receiving high-dose B ion implantation in in situ B-doped SiGe epilayers, then applying ns UV-LA to melt and regrow it [54]. A remaining concern would be crystal defects left after LPER. It is mandatory to melt the ion-implanted region down to the end tail of the dopant profile. Otherwise, the residual point defects and impurities originated from ion implantation precipitate in the form of extended defects in the top of the nonmelted region [30]. Moreover, stacking faults starting from the initial amorphous/crystalline (*a*/*c*) interface can be observed even at such a full melt condition [43]. They might come either from possible nonuniformity of the as-implanted *a*/*c* interface or from nonuniform *l*/*s* interface during UV-LA [55,56].

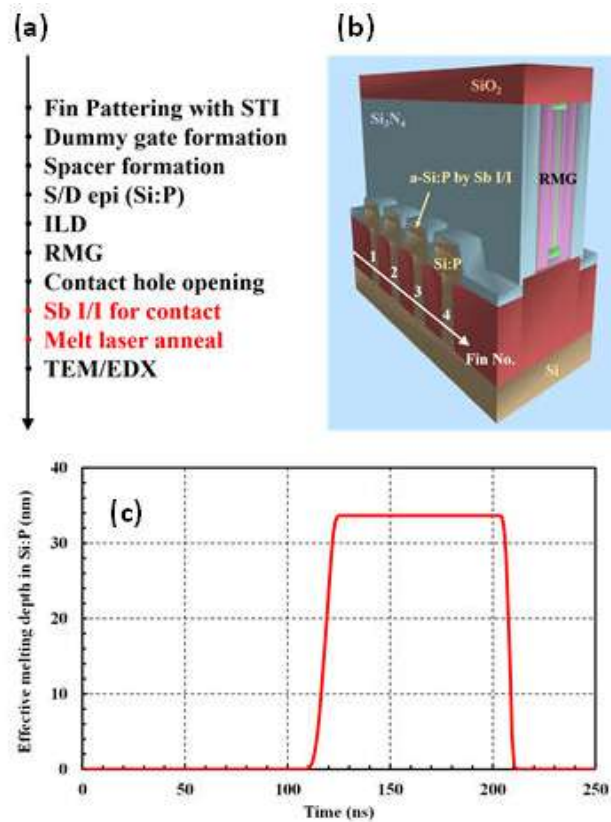

**Figure 9.** *Cont.*

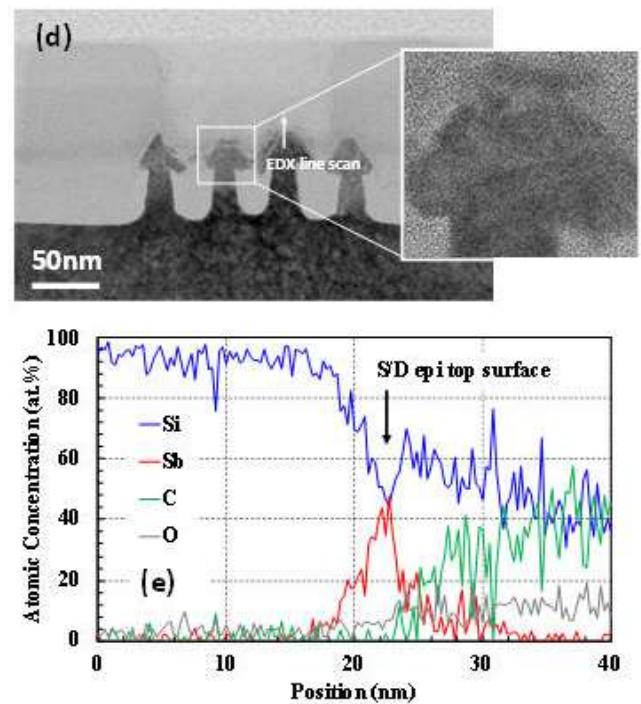

**Figure 9.** (**a**) Process flow of the FinFET contact modules; (**b**) Structure used for the 3D TCAD simulation; (**c**) Simulated time vs. melting Si fin depth profile during ns UV-LA; (**d**) Cross-sectional TEM image of the Si-based contact module after ns UV-LA; (**e**) Energy-dispersive X-ray spectroscopy line scan taken along the annealed Si fin (the carbon signal comes from the structures surrounding the contact hole). Reprinted/adapted under CC BY 4.0 from Ref. [52]. 2020, T. Tabata et al.

### 3.2. Dopant Activation by Solid Phase Epitaxial Regrowth (SPER)

As already mentioned above, another way of enabling the metastable activation of the dopants is SPER. In our previously reported ns UV-LA SPER processes [46], multiplying thermally independent laser shots is necessary to fully crystallize a Si layer amorphized by ion implantation (Figure 10a–c). Then, the moving amorphous/crystalline (*a/c*) interface maintains a good flatness, resulting in a small root-mean-square value of surface roughness (about 0.10 nm) after SPER completion. An extracted maximum crystallization rate is from 0.8 to 1.8 nm per shot in roughly 10 nm thick amorphous Si layers with different dopants such as B, P, and As (Figure 10d). A rough estimation of film resistivity is also provided based on sheet resistance measurements and an assumption that the conducting layer thickness is equal to the amorphization depth. The calculated film resistivity values imply that the active carrier concentration achieved by ns UV-LA SPER could be higher than that of ns UV-LA LPER.

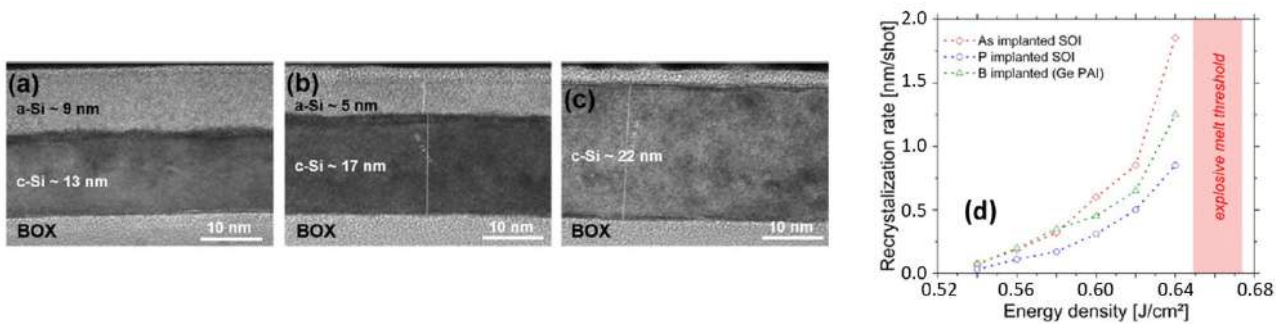

**Figure 10.** (**a**–**c**) Progressive SPER with ns UV-LA multiple pulses in a structure of amorphous Si/SiO$_2$/Si substrate. (**d**) Estimated recrystallization rates for each sample as a function of laser energy density. Reprinted/adapted under CC BY 4.0 from Ref. [46]. 2021, P. Acosta Alba et al.

We have reported similar attempts also for μs UV-LA SPER processes [47,57], where the monocrystalline regrowth and flat surface morphology are basically reproduced as in ns UV-LA ones. Interestingly, as shown in Figure 11a, μs UV-LA SPER processes introduce the surface segregation of dopants (indeed, it is known for furnace SPER of As-implanted Si, as reported in Ref. [58]). It is uncertain if our previously reported ns UV-LA processes also induce it. There might be an impact of the SPER rate, which could be different between the reported ns and μs UV-LA processes because of their different ramp-up and cool-down rates. The active carrier concentration measured by the differential Hall effect methodology (DHEM) [59] outperforms $1 \times 10^{21}$ at./cm$^3$ near the surface. The thermal stability of these active carriers is investigated by using ns UV-LA as deactivation annealing (Figure 11b). The applied deactivation conditions are those which are reported for copper (Cu) or ruthenium (Ru) based the industrial BEOL interconnect annealing by using ns UV-LA. Up to a submillisecond processing, the maximum sheet resistance degradation is limited to about 5%, encouraging UV-LA integration into different stages of the CMOS devices.

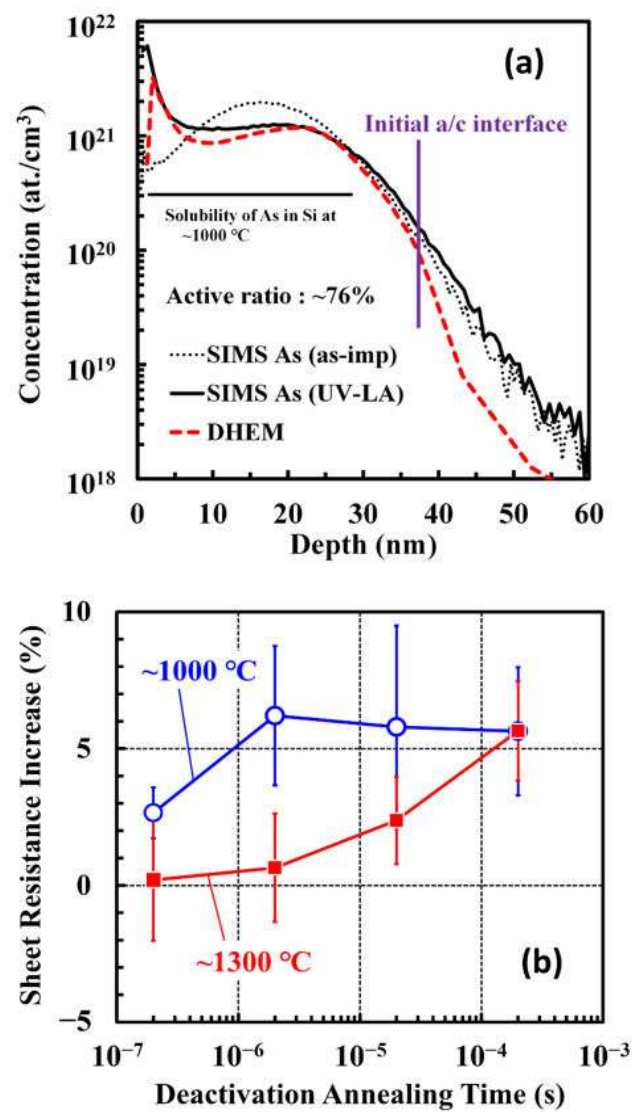

**Figure 11.** (**a**) SIMS and differential Hall effect methodology (DHEM) profiles taken after μs UV-LA SPER. The initial *a/c* interface position and the As solid solubility in *c*-Si at ∼1000 °C reported in Ref. [60] are also indicated. (**b**) Ratio of sheet resistance degradation as a function of the accumulated deactivation annealing time by using the ns UV-LA processes giving different maximum temperatures. Reprinted/adapted under CC BY 4.0 from Ref. [47]. 2022, T. Tabata et al.

## 4. BEOL Applications

The geometry of the BEOL interconnects (i.e., a trench filled with a highly conductive metal) continuously shrinks. As it limits the growth of metallic grains and results in the increasing density of grain boundaries, electron scattering starts to bring a serious demerit in line resistivity. In fact, there is an exponential relationship between the increase of the line resistivity and the scaling of the cross-sectional area of the BEOL lines (Figure 12) [61–64]. Beyond the 7 nm technology node, alternative metals such as Ru [61,63,65] and cobalt (Co) [61,66,67] start to be considered because of their potential benefit in line resistivity. The figure of merit is determined by a complex combination among bulk resistivity, the cross-sectional area of lines, the mean-free-path of electrons, electro-migration reliability (i.e., melting point of metals), integration compatibility (e.g., availability of raw materials, process uniformity), and the impacts from a barrier and liner. However, some efforts to extend the Cu-based BEOL technologies are found [61,68,69], and indeed Cu-based lines can give lower line resistivity than the alternative metals, even in scaled BEOL modules [61,68]. In addition, it is not realistic to replace all Cu lines in the whole BEOL modules, especially for large interconnects (e.g., semiglobal and global lines [70]).

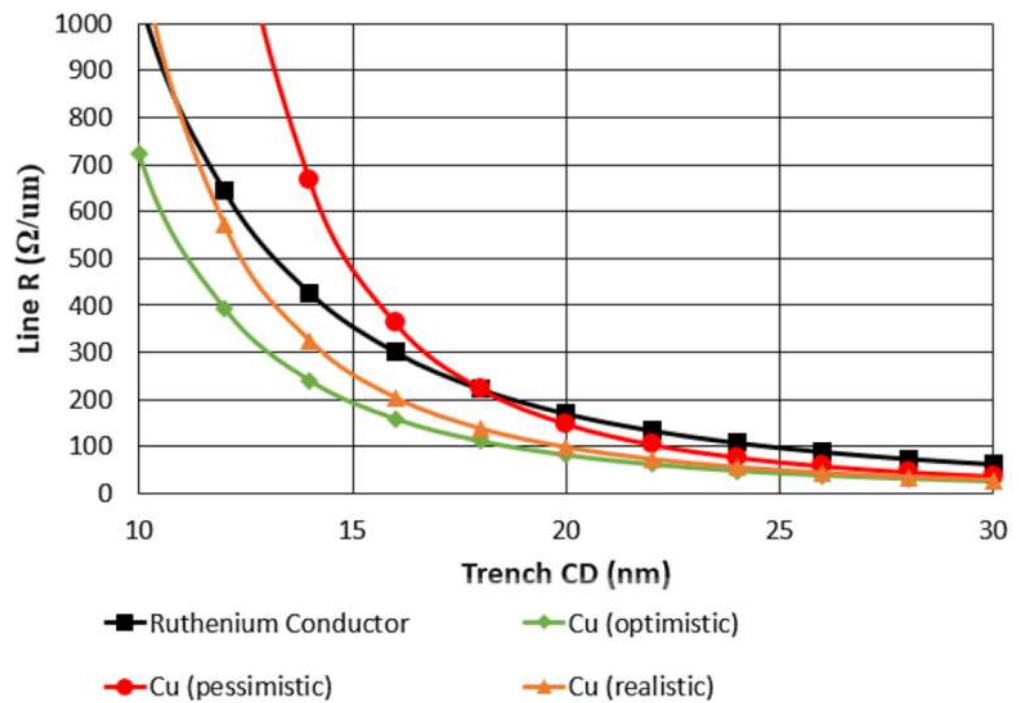

**Figure 12.** Theoretical trends of line resistance vs. line scale (i.e., trench CD) for Cu- and Ru-based interconnects. Reprinted with permission from Ref. [64]. 2020, American Vacuum Society.

In recent years, the impacts of LA have started to be investigated in real BEOL modules or blanket thin films. Its expected benefit is to enable a bamboo-like structure (i.e., reduction of electron scattering spots) in scaled BEOL lines thanks to the capability of processing wafers at a higher temperature (possibly melting Cu but not Ru) than in typical BEOL furnace annealing (e.g., less than 420 °C up to 1 h [71–74]). As shown in Figures 13 and 14, ns LA (non-UV) indeed shows such a potential [75,76].

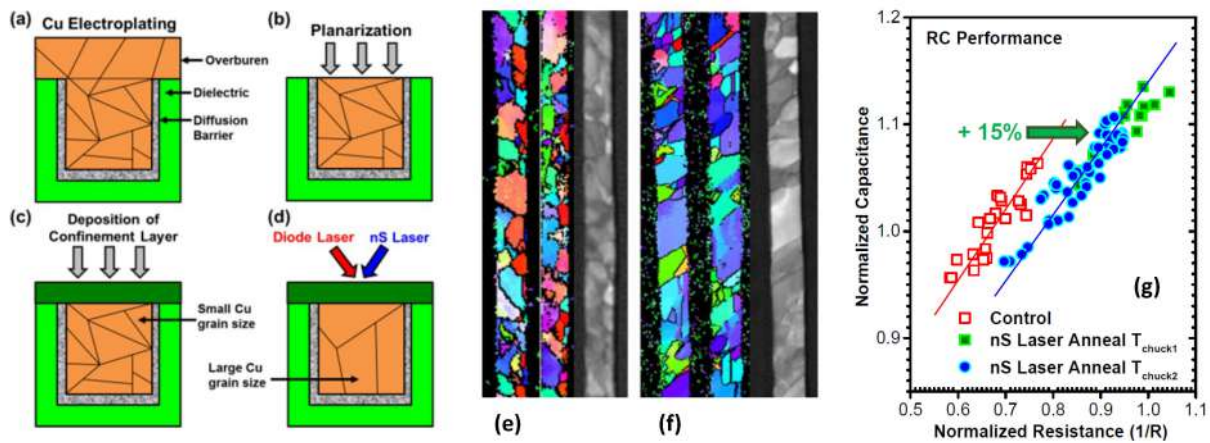

**Figure 13.** (**a**–**d**) Schematics of ns LA process flow, (**e**,**f**) grain analysis performed in the control and Cu melt ns LA lines, (**g**) RC performance measured in the control and Cu melt ns LA lines. Reprinted/adapted with permission from Ref. [75]. 2018, IEEE.

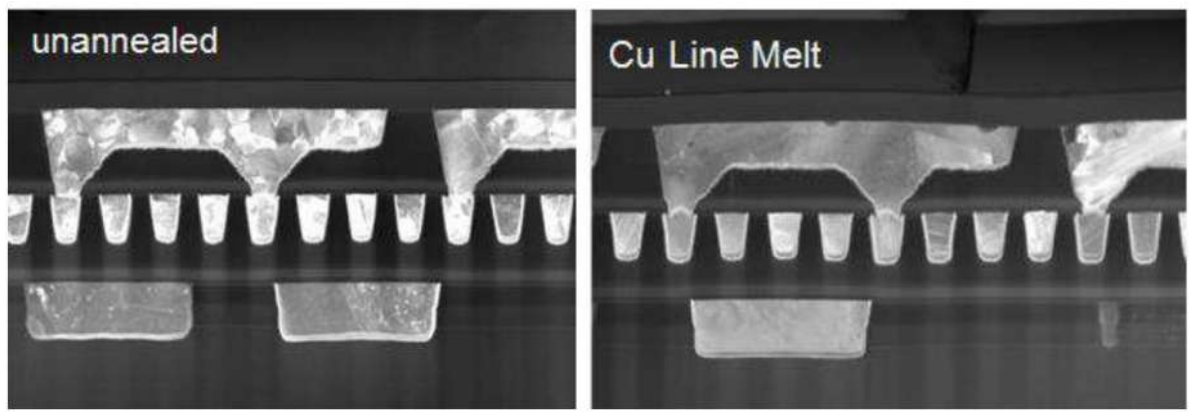

**Figure 14.** Examples of Cu BEOL lines prior to and after Cu melt LA. Reprinted with permission from Ref. [76]. 2018, The Electrochemical Society (ECS).

*4.1. Cu Interconnect*

Unfortunately, there is no published work yet about the use of UV-LA in real Cu BEOL modules. However, we have recently conducted some preliminary works by using blanket Cu thin (roughly 50 nm thick) films. Firstly, it has been demonstrated that µs UV-LA enables much greater grain growth than furnace annealing at both Cu submelt and melt conditions. A furnace process at 600 °C typically gives a mean grain size (Av.) of about 100 nm in 50 nm thick Cu films [77,78], whereas it becomes approximately four times and ten times greater in the Cu submelt and melt µs UV-LA, respectively (Figure 15) [79]. Secondly, it has been revealed that even in such a short timescale of process, the structure (i.e., a typical Cu-based BEOL stack of dielectric/Cu/tantalum (Ta)/SiO$_2$/Si substrate) can be degraded due to interlayer atomic diffusion (Figure 16). Although a process window already exists (e.g., Process A in the Cu submelt shows a 15% reduction of sheet resistance), improving the thermal stability of the Cu-based BEOL structure, for instance, by using an alternative barrier and/or liner (e.g., tantalum nitride (TaN)/Co [68,69], TaN/Ru [68], or tantalum-manganese oxides (TaMn$_x$O$_y$) [68]) would further extend the merit of using UV-LA.

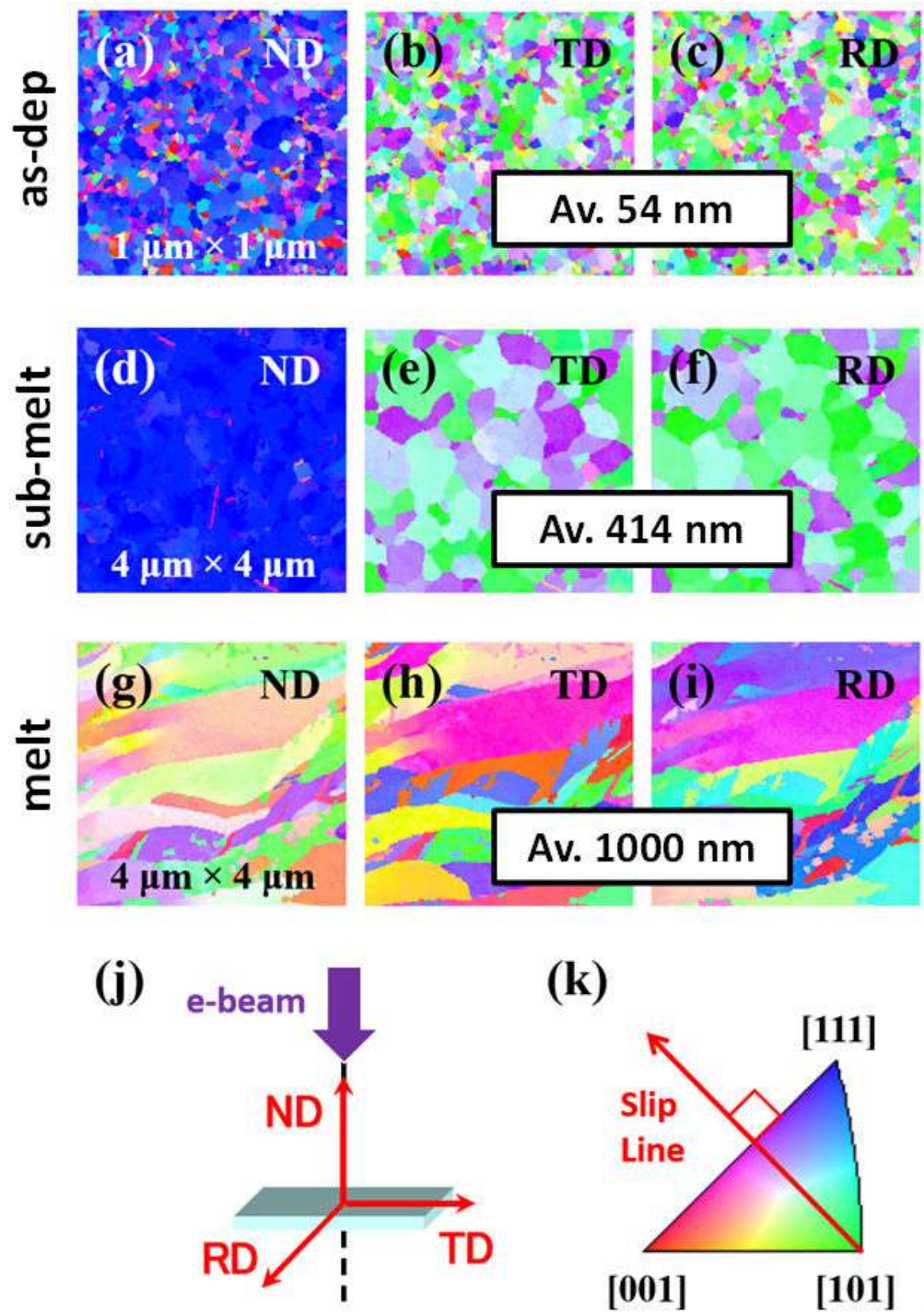

**Figure 15.** Images of electron diffraction mapping taken for the nonannealed (**a**–**c**) and annealed Cu thin films (**d**–**f**) are for a Cu submelt µs UV-LA condition, whereas (**g**–**i**) are for a Cu melt µs UV-LA condition. As depicted in (**j**), ND, TD, and RD stand for normal direction, transverse direction, and reference direction, respectively. Moreover, a standard triangle of grain orientations is shown in (**k**). Reprinted/adapted with permission from Ref. [79]. 2021, IEEE.

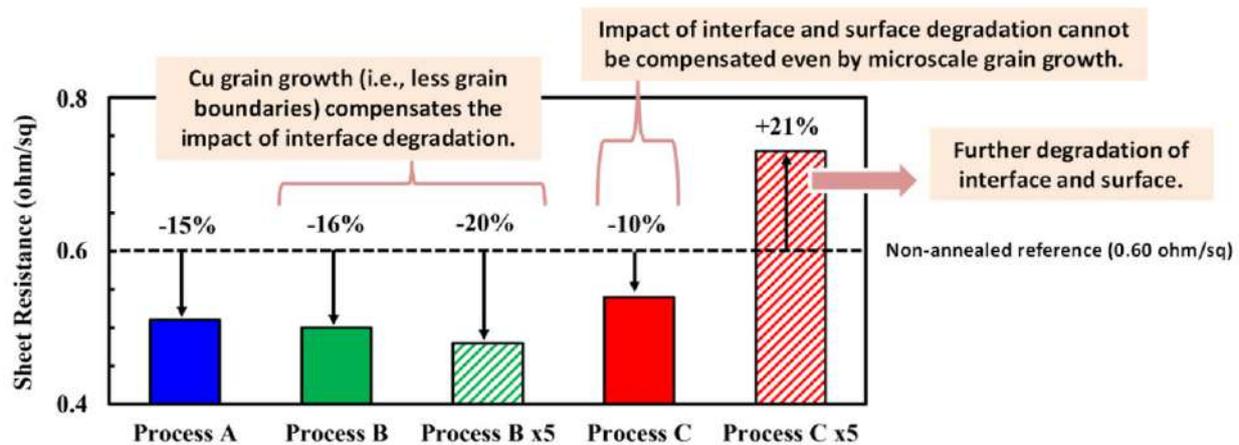

**Figure 16.** Sheet resistance measured in the Cu-based BEOL blanket structure (dielectric/Cu/Ta/SiO$_2$/Si substrate) before and after μs UV-LA (Processes A, B, and C). Reprinted/adapted with permission from Ref. [80]. 2022, IEEE.

*4.2. Ru Interconnect*

Replacement of Cu with an alternative metal in BEOL interconnects is attractive, especially in local lines [70]. One of the listed candidates is Ru, as shown in Figure 17 [81]. Its smaller mean free path of electrons ($\lambda$) than Cu reduces electron scattering and compensates a disadvantage of bulk resistivity ($\rho_0$). Moreover, its higher melting point than Cu is expected to improve electro-migration performance. Furthermore, Ru is already used in industry-like Cu-based BEOL interconnects as a liner material [82]. Although it is not a focus of this review article, Ru-based BEOL processes (e.g., deposition, CMP, and etching) could therefore rapidly mature.

We have previously evaluated the impact of ns UV-LA on the line resistance of Ru lines fabricated by a 21 nm half-pitch dual-damascene process [83]. Then, although the Ru lines are not supposed to be melted because of its much higher (more than two times) melting point than Cu, the line resistance is reduced for up to 25% by multiple ns UV-LA processes compared to the as-deposited Ru lines.

On the other hand, in a way of further scaling in BEOL interconnects, semidamascene processing (also called "subtractive") is emerging for Ru-based BEOL interconnects [84,85]. This new approach might be more adapted for UV-LA because it allows to anneal the metal before line patterning, and thereby may alleviate the challenge of controlling the effects of light pattern interference. In this context, we have also evaluated grain growth in deposited Ru thin films having a similar thickness to the Cu studies [74,83,86], demonstrating effective grain enlargement and the associated sheet resistance drops with multiple ns UV-LA processes.

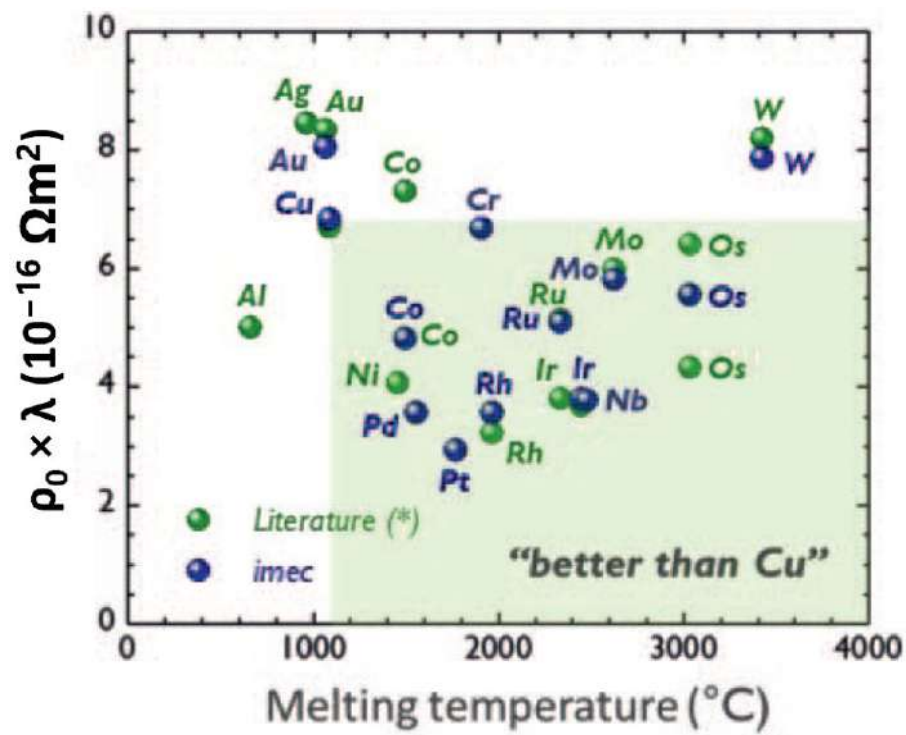

**Figure 17.** Figure of merit defined as mean free path ($\lambda$) × bulk resistivity ($\rho_0$) vs. melting point for different metals. Reprinted/adjusted with permission from Ref. [81]. 2018, IEEE.

## 5. "More Than Moore" Applications

Not only in 3D-integrated devices but also in planar ones is the BEOL module the place where the applicable thermal budget is strongly restricted. It often makes material engineering difficult and narrows the range of applications. However, if this restriction is removed, then the diversification of applications will start like cutting a weir. We hereafter present a couple of examples of material engineering in BEOL allowed by short timescale UV-LA.

*5.1. Large Poly-Si Grain Formation from Amorphous Si Thin Film*

The formation of an electrically active layer (i.e., doped semiconductor with a monocrystalline or polycrystalline state) in BEOL may open a chance for disruptive innovation to boost CMOS performance. However, the maximum process temperature in BEOL is generally much lower than the one enabling solid phase crystallization (SPC) of amorphous Si on $SiO_2$ (e.g., 600 °C for several hours [87,88]). Therefore, there is a clear need for utilizing short timescale UV-LA, with which both liquid phase crystallization (LPC) and SPC will become accessible.

Recently, FinFET device integration into BEOL has been demonstrated with the aim of reducing chip size and power consumption [2]. In this work, a LA (non-UV) process is performed as a key process for crystallizing amorphous Si via LPC. Such a potential has been also demonstrated by ns UV-LA, as shown in Figure 18 [1]. Si-based photonic device integration in BEOL via ns UV-LA LPC can be also found in the literature [89].

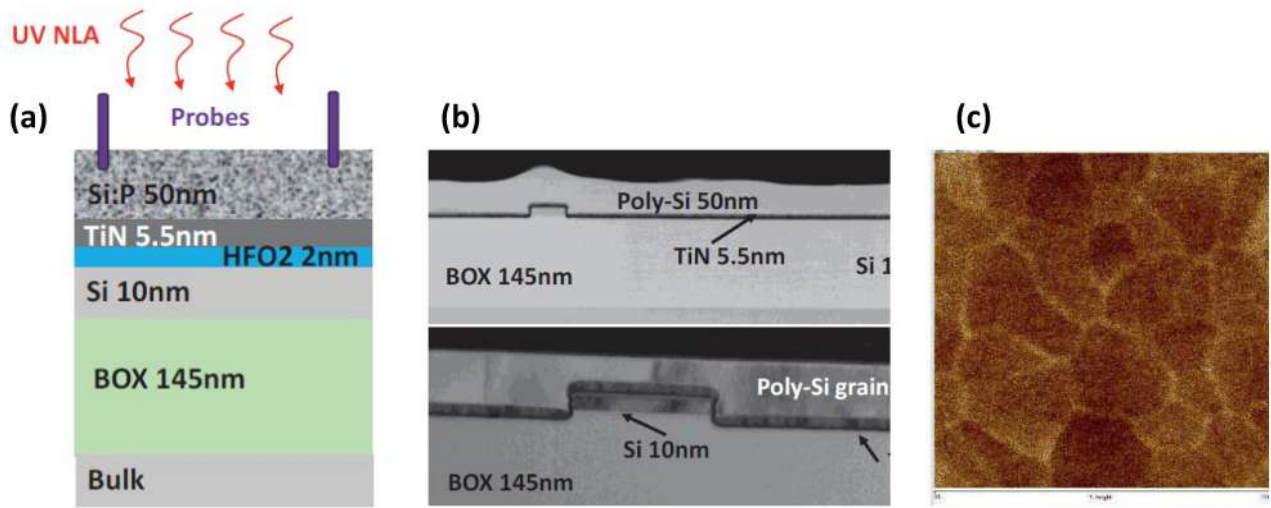

**Figure 18.** (**a**) Schematic of ns UV-LA in a 3D type structure having an amorphous Si layer on the top, (**b**) cross-sectional TEM images taken in the test structure followed by ns UV-LA and top amorphous Si melting then by CMP to planarize the top surface, (**c**) atomic force microprobe image taken after the CMP planarization. Reprinted/adjusted with permission from Ref. [1]. 2018, IEEE.

It should be noted that melting materials in a device structure may engender drawbacks in subsequent processes. As already shown in Figure 18b, Si LPC clearly presents hillocks on the just-crystallized poly-Si surface, and they must be planarized by CMP to follow the device fabrication process flow. It occurs due to the collision of the solidification front among grains crystallizing from the liquid Si, and the hillocks are formed on the final grain boundaries [90]. Interestingly, in the case of the LPER (i.e., no poly grain boundaries) in Si channel extension, shown in Figure 4d, the regrown Si surface keeps a good flatness and does not impact the subsequent S/D epitaxy. Although the surface tension [91] induced on the top of the liquid Si may induce roughening of the regrown surface [92], a thick (tens of nanometers) dielectric capping layer deposited on the partially amorphized Si channel seems to help to avoid the degradation of the surface morphology. Therefore, one may suppose that a similar capping approach could work also for the Si LPC. However, careful consideration would be necessary, because it is known that a periodic surface pattern (called "wrinkles") emerges and evolves in a stack having an elastic layer on a liquid substrate [93]. Our recent study has revealed that the Si melt induced by ns UV-LA in $SiO_2$/Si stacks indeed triggers the formation of wrinkles and they grow in a spatial wavelength and height as the maximum melted Si depth increases [94]. As this phenomenon is described by dimensional, mechanical, and viscoelastic parameters of the materials involved in the system, the stack to be treated by ns UV-LA should be carefully designed.

*5.2. Hf-Based Ferroelectric Layer Formation*

Another application in BEOL which a short timescale UV-LA may enable is the formation of a high relative permittivity or ferroelectric layer (e.g., capacitors integrated into 130 nm CMOS BEOL in Ref. [12]). Hf-based oxides are today well-known as a CMOS technology compatible with ferroelectric thin films [12,95,96]. Their dielectric functionalities (i.e., high relative permittivity or ferroelectricity) rely on crystal phases, and cubic (*c*), tetragonal (*t*), orthorhombic (*o*), and monoclinic (*m*) ones are typical. The ferroelectricity of $HfO_2$ thin films is related to the *o*-phase (e.g., $Pca2_1$ space group), which emerges during the transition from the *t*-phase to the *m*-phase [97,98]. Therefore, it is critical to control the kinetics of nucleation and crystal phase transformation.

Inspired by previous studies [99,100], we have come up with an idea of shedding a light on it by exploring different process timescales (i.e., dwell time) with ns [101] and μs [102] UV-LA. The expected impact of controlling the dwell time is depicted in Figure 19a. Then, the relative permittivity of the annealed $HfO_2$ thin film evolves with μs-scale dwell times, as shown in Figure 19b, supporting the proposed idea. The appearance of ferroelectricity is also experimentally evidenced in this study by means of PUND (Positive Up Negative Down)-corrected polarization–voltage (P-V) measurements [103]. When squeezing the dwell timescale toward the ns range, it is expected that $HfO_2$ phase control comes to rely on nucleation rather than phase transformation. Indeed, repetition of the ns UV-LA process results in a cumulative crystallization of the same phase, as shown in Figure 20. The selectivity of the phase to nucleate in the ns UV-LA seems dependent on the maximum process temperature [12], doping content [12], and stack [101]. As a temperature range of interest is the one near or beyond the Si melting point [12], it is necessary to insert additional layers for efficient UV laser absorption and heat confinement. An example is a stack of metal/$HfO_2$/metal/$SiO_2$/Si, as shown in Ref. [12], where the metal layers work for UV laser absorption and the $SiO_2$ layer does for heat confinement in the upper metal/$HfO_2$/metal part. Silicon (amorphous, polycrystalline, or monocrystalline) can also be used instead of the metal layers. Some doping elements are known to stabilize the ferroelectric phase in $HfO_2$ [104,105]. At a given process temperature, the cooling profile may be slightly modulated by the stack and allow partial phase transformation within the nuclei (but their growth with a single ns laser shot would be almost negligible because of limited atomic diffusion in the ns-scale [101]).

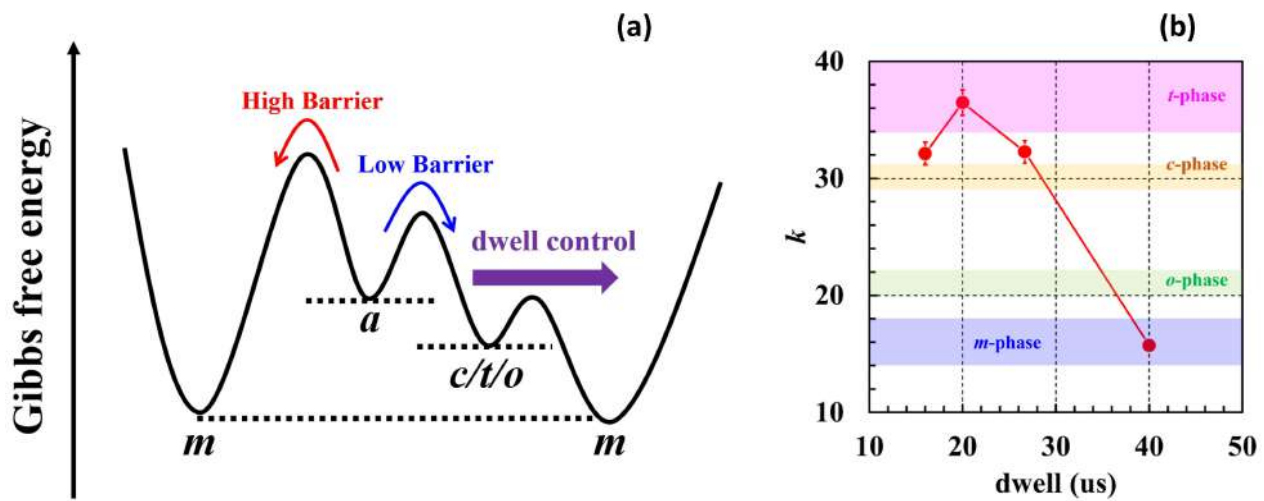

**Figure 19.** (**a**) Schematic Gibbs free energy diagram of different $HfO_2$ crystal phases, where the character "*a*" stands for the amorphous state, "*c*" for the *c*-phase, "*t*" for the *t*-phase, "*o*" for the *o*-phase, and "*m*" for the *m*-phase, respectively. (**b**) Relative permittivity (*k*) values of the 10 nm thick $HfO_2$ films as a function of the UV-LA dwell time. The theoretically predicted *k*-value range of each phase is shown together. Reprinted/adjusted with permission from Ref. [102]. 2021, The Japan Society of Applied Physics.

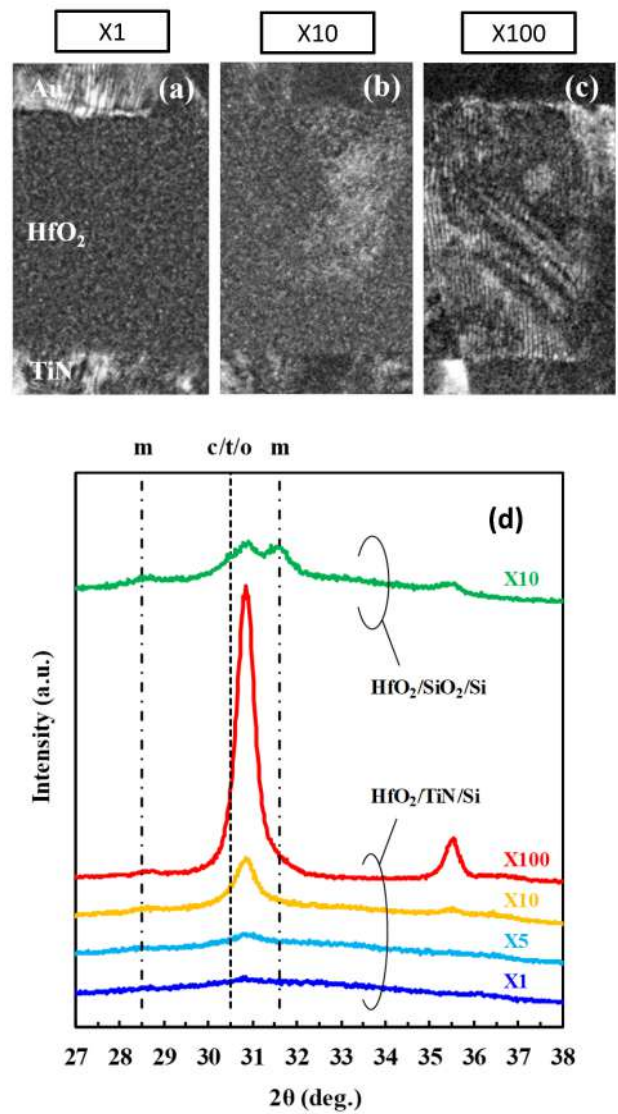

**Figure 20.** (**a**–**c**) Cross-sectional dark-field TEM images and (**d**) XRD patterns taken on 50 nm thick HfO$_2$/TiN/Si samples after single or multiple ns UV-LA processing. Reprinted/adjusted with permission from Ref. [101]. 2020, The Japan Society of Applied Physics.

## 6. Conclusions

In this review, recent progresses of ns and μs UV-LA technologies have been reviewed, focusing on applications relevant to CMOS devices. The selectivity of heating in the vertical direction that short timescale UV-LA provides is highlighted as a key feature for 3D integration. The presented applications vary from FEOL to BEOL, indicating the high potential of integrating UV-LA processes into different stages of a CMOS fabrication flow. From the viewpoint of materials science, short timescale UV-LA opens new fields of research, especially related to its nonequilibrium aspect. Although the reported range of process timescale from ns to μs is long enough to be at thermal equilibrium [106], it is still chemically and mechanically out of equilibrium. Therefore, phenomena such as atomic bonding rearrangement, diffusion, activation, crystallization (including grain growth), and stress relaxation show interesting behaviors. From them, module-level major improvements to boost the entire CMOS performance may be achieved. Even if there is still a lot of effort required to integrate the presented UV-LA processes into the industrial CMOS devices, the technological maturity of short-timescale UV-LA is steadily progressing.

**Author Contributions:** Writing—original draft preparation, T.T.; writing—review and editing, F.R., L.T., S.H., P.-E.R., I.K. and K.H. All authors have read and agreed to the published version of the manuscript.

**Funding:** This project has received funding from the ECSEL Joint Undertaking (JU) under grant agreement no. 875999. The JU receives support from the European Union's Horizon 2020 research and innovation programme, and the Netherlands, Belgium, Germany, France, Austria, Hungary, United Kingdom, Romania, and Israel.

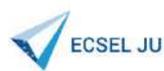 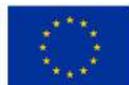

**Institutional Review Board Statement:** Not applicable.

**Informed Consent Statement:** Not applicable.

**Data Availability Statement:** Not applicable.

**Acknowledgments:** The authors thank their former colleagues, Joris Aubin and Fulvio Mazzamuto, for their contribution to the works cited in this review article, as well as all other LASSE employees for their contribution to the development of ns and µs UV-LA platforms.

**Conflicts of Interest:** The authors declare no conflict of interest.